\documentstyle[12pt,epsfig]{article}
\newcommand{\nd}{\noindent}
\newcommand{\beq}{\begin{equation}}
\newcommand{\eeq}{\end{equation}}
\newcommand{\barr}{\begin{eqnarray}}
\newcommand{\earr}{\end{eqnarray}}
\newcommand{\ba}{\begin{array}}
\newcommand{\ea}{\end{array}}
\newcommand{\bfp}{\mbox{\boldmath $p$}}
\newcommand{\bfk}{\mbox{\boldmath $k$}}

\newcommand{\bfz}{\mbox{\boldmath $0$}}

\newcommand{\la}{\lambda}

\newcommand{\pup}{p^\uparrow}
\newcommand{\aup}{a^\uparrow}
\newcommand{\pdown}{p^\downarrow}
\newcommand{\adown}{a^\downarrow}
\newcommand{\NP}[1]{{\it Nucl.\ Phys.}\ {\bf #1}}
\newcommand{\ZP}[1]{{\it Z.\ Phys.}\ {\bf #1}}
\newcommand{\PL}[1]{{\it Phys.\ Lett.}\ {\bf #1}}
\newcommand{\PR}[1]{{\it Phys.\ Rev.}\ {\bf #1}}
\newcommand{\PRL}[1]{{\it Phys.\ Rev.\ Lett.}\ {\bf #1}}

\def\lsim{\mathrel{\rlap{\lower4pt\hbox{\hskip1pt$\sim$}}\raise1pt\hbox{$<$}}}
\def\gsim{\mathrel{\rlap{\lower4pt\hbox{\hskip1pt$\sim$}}\raise1pt\hbox{$>$}}}
\def\nostrocostruttino#1\over#2{\mathrel{\mathop{\kern 0pt \rlap
{\hbox{$#1$}}} \hbox{\kern-.135em $#2$}}}
\def\sumint{\nostrocostruttino \sum \over {\displaystyle\int}}
\begin{document}
\begin{flushright}
DFTT 04/99 \\
VUTH 99-02 \\
INFNCA-TH9902 \\
hep-ph/9901442 \\
\end{flushright}
\vskip 1.5cm
\begin{center}
{\bf Phenomenology of single spin asymmetries in \mbox{\boldmath{$p^{\uparrow} \!\!p \to 
\pi X$}} }\\
\vskip 0.8cm
{\sf M.\ Anselmino$^1$, M.\ Boglione$^2$, F.\ Murgia$^3$}
\vskip 0.5cm
{$^1$ Dipartimento di Fisica Teorica, Universit\`a di Torino and \\
      INFN, Sezione di Torino, Via P. Giuria 1, 10125 Torino, Italy\\
\vskip 0.5cm
$^2$  Dept. of Physics and Astronomy, Vrije Universiteit Amsterdam, \\
De Boelelaan 1081, 1081 HV Amsterdam, The Netherlands \\ 
\vskip 0.5cm
$^3$  Istituto Nazionale di Fisica Nucleare, Sezione di Cagliari\\
      and Dipartimento di Fisica, Universit\`a di Cagliari\\
      C.P. 170, I-09042 Monserrato (CA), Italy} \\
\end{center}
\vskip 1.5cm
\noindent
{\bf Abstract:} \\ 
A phenomenological description of single transverse spin effects in 
hadron-hadron inclusive processes is proposed, assuming 
a generalized factorization scheme and pQCD hard interactions. 
The transverse momentum, $\bfk_\perp$, 
of the quarks inside the hadrons and of the hadrons relatively to the
fragmenting quark, is taken into account in distribution and fragmentation 
functions, and leads to possible non zero single spin asymmetries. 
The role of $\bfk_\perp$ and spin dependent quark fragmentations -- the 
so-called Collins effect -- is investigated in details in $p^{\uparrow}p 
\to \pi X$ processes: it is shown how the experimental data could 
be described, obtaining an explicit expression for the
spin asymmetry of a polarized fragmenting quark, on which some comments
are made. Predictions for other processes, possible further applications 
and experimental tests are discussed. 

\newpage
\pagestyle{plain}
\setcounter{page}{1}
\nd
{\bf 1. Introduction}
\vskip 6pt
Understanding the kinematics and the dynamics of the quarks inside the hadrons 
is a challenging and interesting problem. In simple quark-parton models 
the quarks inside a fast hadron are assumed to be collinear, {\it i.e.} their 
momentum is parallel to the momentum of the hadron. 
This assumption works well for unpolarized processes, like $A+B \to C+X$
which are computed, at large energy and momentum transfer $p_T$, 
by the usual convolution of distribution and fragmentation functions
with a hard perturbative elementary interaction, according to QCD 
factorization theorem \cite{cms}. However, in polarized reactions, the 
contribution of transverse intrinsic momenta often turns out to be 
crucial in understanding the experimental results. 

In Ref. \cite{noi1}, we proved how the large single spin asymmetries 
\beq
A_N(x_F, p_T) = \frac{ d\sigma^{\uparrow} - d\sigma^{\downarrow} }
           { d\sigma^{\uparrow} + d\sigma^{\downarrow} }
    = \frac{ d\sigma^{\uparrow} - d\sigma^{\downarrow} }
           { 2 \, d\sigma^{unp} } \>, \label{an}
\eeq
measured in $p^{\uparrow}p \to \pi X$ by the E704 Group at Fermilab
\cite{e704}, can be accounted for by introducing \bfk$_{\perp}$ and spin 
dependences in the distribution functions of quarks inside the initial 
polarized proton. 
$d\sigma^{\uparrow,\downarrow}$ stands for the differential cross section 
$E_\pi \, d^3\sigma^{p^{\uparrow,\downarrow}p \to \pi X} /d^3
\mbox{\boldmath $p$}_{\pi}$; 
$x_F = 2p_L/\sqrt{s}$, where $p_L$ is the pion longitudinal momentum in the 
$p\,p$ c.m. frame, and $\sqrt{s}$ is the total c.m. energy. 
The proton spin is ``up" or ``down" with respect to
the scattering plane: $\uparrow$ ($\downarrow$) is the direction parallel
(antiparallel) to $\bfp_{\pup} \times \bfp_\pi$.
Other sizeable single spin asymmetries have been explained
or predicted in Ref. \cite{noi2}. 

The original suggestion that the intrinsic $\bfk_\perp$ of the quarks in
the distribution functions might give origin to single spin asymmetries
was first made by Sivers \cite{siv}; such an effect is not forbidden
by QCD time reversal invariance \cite{col} provided one takes into account 
soft initial state interactions among the colliding hadrons \cite{noi1}. 
A similar suggestion for possible origin of single spin asymmetries
was later made by Collins \cite{col}, concerning 
transverse momentum effects in the fragmentation of a polarized quark. 
A first simple application of this idea was given in \cite{art}.    

Qiu and Sterman \cite{qs1, qs2} have proven that a generalized 
factorization theorem holds in QCD with twist-3 distribution 
and/or fragmentation functions, which take into account initial or final 
state interactions. Recently \cite{qs3} they have computed - using a simple 
model for a new twist-3 correlation function -- single spin asymmetries for 
$p^{\uparrow}p \to \pi X$ and $\bar p^{\uparrow}p \to \pi X$ processes.
Their approach establishes a sound theoretical ground for simple 
phenomenological applications: a new quantity is introduced for valence
$u$ and $d$ quarks which represents a correlation between these quarks
inside a polarized proton and an external gluonic field; a simple 
parametrization of the new function is fixed by fitting data and some
predictions for other observables are given. 

In this respect the computation of Ref. \cite{qs3} is similar to that
of Ref. \cite{noi1}: also in the latter a new quantity was introduced
for valence $u$ and $d$ quarks, a quantity which requires some initial state
interactions between the colliding hadrons in order not to be forbidden
by time reversal invariance, and the assumption that a factorization 
scheme still holds; a simple parametrization was given and 
fixed by fitting the experimental data, and predictions were subsequently 
made for other processes \cite{noi2}. 

The main difference is that in Sivers and our approach \cite{noi1,noi2,siv} 
-- based on the QCD improved parton model -- the new quantity,
called $\Delta^N f_{a/\pup}(x_a, \bfk_{\perp a})$ or 
$2I^{a/p}_{+-}(x_a, \bfk_{\perp a})$, has a partonic interpretation:
it is the difference between the density numbers 
$\hat f_{a/\pup}(x_a, \bfk_{\perp a})$ and 
$\hat f_{a/\pdown}(x_a, \bfk_{\perp a})$
of partons $a$, with all possible polarization,
longitudinal momentum fraction $x_a$ and intrinsic transverse momentum 
$\bfk_{\perp a}$, inside a transversely polarized proton with spin $\uparrow$ 
or $\downarrow$: 
\barr
\Delta^Nf_{a/\pup}(x_a, \bfk_{\perp a})  &\equiv& 
\hat f_{a/\pup}(x_a, \bfk_{\perp a})-\hat f_{a/\pdown}(x_a, \bfk_{\perp a}) 
\label{del1}\\
& = & \hat f_{a/\pup}(x_a, \bfk_{\perp a})-\hat f_{a/\pup}(x_a, - 
\bfk_{\perp a}) \nonumber
\earr
where the second line follows from the first one by rotational invariance.
Notice that $\Delta^Nf_{a/\pup}(x_a,\bfk_{\perp a})$ 
vanishes when $\bfk_{\perp a} \to 0$; parity invariance also requires 
$\Delta^Nf$ to vanish when the proton transverse spin has no component
perpendicular to $\bfk_{\perp a}$, so that
\beq
\Delta^Nf_{a/\pup}(x_a,\bfk_{\perp a}) 
\sim k_{\perp a} \, \sin\alpha
\label{sind}
\eeq
where $\alpha$ is the angle between $\bfk_{\perp a}$ and the $\uparrow$ 
direction.

$\Delta^N f$ by itself is then a leading twist distribution function, but
its $\bfk_\perp$ dependence, when convoluted with the 
elementary partonic cross-section, results in twist-3 contributions
to single spin asymmetries. This same function (up to some factors)
has also been introduced in Ref. \cite{mul1} -- where it is 
denoted by $f_{1T}^\perp$ -- as a leading twist $T$-odd distribution function.
In Ref. \cite{bmt} the relation between $f_{1T}^\perp$ and twist-3
correlation functions is explained and an evaluation of single spin asymmetries
in Drell-Yan processes, originating from a gluonic background and the
so-called gluonic poles, is given. The exact relation between 
$\Delta^N f$ and $f_{1T}^\perp$ is discussed in Ref. \cite{bm}.

A function analogous to $\Delta^Nf_{a/\pup}(x_a, \bfk_{\perp a})$
can be defined for the fragmentation process of a transversely 
polarized parton \cite{col}, giving the difference between the 
density numbers $\hat D_{h/\aup}(z, \bfk_{\perp h})$ and 
$\hat D_{h/\adown}(z, \bfk_{\perp h})$
of hadrons $h$, with longitudinal momentum fraction $z$ and transverse 
momentum $\bfk_{\perp h}$ inside a jet originated by the fragmentation of a 
transversely polarized parton with spin $\uparrow$ or 
$\downarrow$:
\barr
\Delta^N D_{h/\aup}(z, \bfk_{\perp h}) &\equiv&
\hat D_{h/\aup}(z, \bfk_{\perp h}) - \hat D_{h/\adown}(z, \bfk_{\perp h}) 
\label{delf1}\\
&=& \hat D_{h/\aup}(z, \bfk_{\perp h})-\hat D_{h/\aup}(z, - \bfk_{\perp h}) \>.
\nonumber
\earr
A closely related function is denoted by $H_1^\perp$ in 
Refs. \cite{mul1, mul2} and its correspondence with $\Delta^ND$ 
is discussed in Ref. \cite{bm}. Again we expect
\beq
\Delta^ND_{h/\aup}(z,\bfk_{\perp h}) 
\sim k_{\perp h} \, \sin\beta
\label{sinf}
\eeq
where $\beta$ is the angle between $\bfk_{\perp h}$ and the $\uparrow$ 
direction.

We first discuss here the role of the above functions, (\ref{del1}) and 
(\ref{delf1}), in generating single spin asymmetries in the large $p_T$
inclusive production of particles, $ A^\uparrow + B \to C + X$, within
a phenomenological QCD factorization model; we give explicit expressions for 
the single spin asymmetries taking into account the leading order 
contributions of parton transverse momentum (Section 2).
In Section 3 we assume that only Collins effect is active -- the analogous
work for Sivers effect was performed in Refs. \cite{noi1, noi2} -- and show 
that it can explain existing data on $\pup p \to \pi X$, with some possible
problems at the largest $x_F$ values; we obtain
an explicit expression for the function given in Eq. (\ref{delf1}). 
Such function is then used to predict other single spin asymmetries in
Section 4, while further comments are made and future applications, with
attention to planned experiments, are discussed in Section 5.

\goodbreak
\vskip 12pt
\nd
{\bf 2. Single spin asymmetries in QCD parton model}
\vskip 6pt

We  write the differential cross-section for
the hard scattering of a polarized hadron $A^\uparrow$ off an unpolarized
hadron $B$, resulting in the inclusive production of a hadron $C$ with
energy $E_C$ and three-momentum $\bfp_C$, $A^\uparrow + B \to C + X$, 
in a factorized form, as 
\barr
d\sigma^\uparrow &\equiv&
\frac{E_C \, d\sigma^{A^\uparrow B \to C X}} {d^{3} \bfp_C} \nonumber \\
&=& {1\over 2} \sum_{a,b,c,d} \>
\sum_{\la^{\,}_a, \la^{\prime}_a, \la^{\,}_b, \la^{\,}_c, \la^{\prime}_c,
\la^{\,}_d, \la^{\,}_C}
\int \frac{dx_a \, dx_b}{\pi z\;16 \pi \hat s ^{2}} \;
 d^2 \bfk_{\perp a} \, d^2 \bfk_{\perp b} \, d^2 \bfk_{\perp C} 
\label{sigmaup} \\
& & \rho_{\la^{\,}_a, \la^{\prime}_a}^{a/A^\uparrow} \,
\hat f_{a/A^\uparrow}(x_a,\bfk _{\perp a}) \, 
\hat f_{b/B}(x_b,\bfk _{\perp b}) \,
\hat M_{\la^{\,}_c, \la^{\,}_d; \la^{\,}_a, \la^{\,}_b} \,
\hat M^*_{\la^{\prime}_c, \la^{\,}_d; \la^{\prime}_a, \la^{\,}_b} \,
\hat D_{\la^{\,}_C,\la^{\,}_C}^{\la^{\,}_c,\la^{\prime}_c}(z,\bfk_{\perp C})
\>. \nonumber
\earr
If we choose the $z$-axis as the direction of motion of $A^\uparrow$
and $xz$ as the scattering plane, then the $\uparrow$ direction is
along the $y$-axis.  

The above expression is proven for collinear configurations, according to 
the QCD factorization theorem: we assume here its validity also 
when taking into account the parton transverse motion. There is no
demonstration of the factorization theorem in such a case, and Eq. 
(\ref{sigmaup}) has to be considered as a plausible phenomenological model.

In our expression (\ref{sigmaup}) we have explicitely shown the $\bfk_\perp$
dependence in the parton distribution and fragmentation. 
The usual number density of partons $a$ inside hadron $A$ is given by
$$
f_{a/A}(x_a) = \int d^2 \bfk_{\perp a}\,\hat f_{a/A^\uparrow}(x_a,
\bfk_{\perp a})\>.
$$
Notice that while $f_{a/A}(x_a)$ cannot depend on the parent hadron spin, 
$\hat f_{a/A^\uparrow}(x_a, \bfk_{\perp a})$ can, provided some 
soft initial state interactions are taken into account \cite{noi1, siv}.
 
$\rho_{\la^{\,}_a, \la^{\prime}_a}^{a/A^\uparrow}(x_a)$ is the helicity 
density matrix of parton $a$ inside the polarized hadron $A$.
The $\hat M_{\la^{\,}_c, \la^{\,}_d; \la^{\,}_a, \la^{\,}_b}$'s
are the helicity amplitudes for the elementary process $ab \to cd$;
if one wishes to consider higher order (in $\alpha_s$) contributions also
elementary processes involving more partons should be included. 
$\hat D_{\la^{\,}_C,\la^{\prime}_C}^{\la^{\,}_c,\la^{\prime}_c}(z, 
\bfk_{\perp C})$ is the 
product of {\it fragmentation amplitudes} for the $c \to C + X$ process
\beq
\hat D_{\la^{\,}_C,\la^{\prime}_C}^{\la^{\,}_c,\la^{\prime}_c} 
= \> \sumint_{X, \la_{X}} {\hat{\cal D}}_{\la^{\,}_{X},\la^{\,}_C;
\la^{\,}_c} \, {\hat{\cal D}}^*_{\la^{\,}_{X},\la^{\prime}_C; \la^{\prime}_c}
\, ,
\label{framp}
\eeq
where the $\sumint_{X, \la_{X}}$ stands for a spin sum and phase space
integration of the undetected particles, considered as a system $X$.
The usual unpolarized fragmentation function $D_{C/c}(z)$, {\it i.e.} 
the density number of hadrons $C$ resulting from the fragmentation of 
an unpolarized parton $c$ and carrying a fraction $z$ of its momentum,
is given by
\beq
D_{C/c}(z) = {1\over 2} \sum_{\la^{\,}_c,\la^{\,}_C} \int d^2 \bfk_{\perp C} 
\, \hat D_{\la^{\,}_C,\la^{\,}_C}^{\la^{\,}_c,\la^{\,}_c}(z, \bfk_{\perp C})
\,. \label{fr}
\eeq

We shall neglect in Eq. (\ref{sigmaup}), due to the limited $Q^2$ range 
of its application, the (unknown) $Q^2$ 
scale dependences in $\hat f$ and $\hat D$; the variable $z$ is related to 
$x_a$ and $x_b$ by the usual imposition of energy momentum conservation
in the elementary 2 $\to$ 2 process, which reads, in the collinear case,
$z = -(x_a t + x_b u)/x_a x_b s$, where $s,t,u$ 
are the Mandelstam variables for the overall process $AB \to CX$, whereas 
$\hat s, \hat t, \hat u$ are those for the elementary process $ab \to cd$. 
A similar expression holds when taking $\bfk_\perp$ into account.
The elementary process amplitudes are normalized so that 
\beq
\frac{d\hat\sigma}{d\hat t} = \frac{1}{16\pi\hat s^2}\frac{1}{4}
\sum_{\la^{\,}_a, \la^{\,}_b, \la^{\,}_c, \la^{\,}_d}
|\hat M_{\la^{\,}_c, \la^{\,}_d; \la^{\,}_a, \la^{\,}_b}|^2\,.
\label{norm}
\eeq

The cross-section $d\sigma^\downarrow$ is readily obtained from 
Eq. (\ref{sigmaup}) by changing $\uparrow$ into $\downarrow$; in absence of
$\bfk_\perp$, collinear configurations and helicity conservation in the
pQCD elementary processes do not allow any single hadron spin dependence  
and it would result $d\sigma^\uparrow = d\sigma^\downarrow$ \cite{noi3}.
With non zero $\bfk_\perp$ instead, spin dependences might still remain
in $\hat f_{a/A^\uparrow}(x_a, \bfk_{\perp a})$ and/or in 
the fragmentation process of a polarized quark.

We specialize now Eq. (\ref{sigmaup}) to the case of $\pup p \to \pi X$
processes; we keep only the leading contributions in $\bfk_\perp$, {\it i.e.}
we consider parton transverse motion only in those functions which
would otherwise be zero [$\Delta^Nf_{a/\pup}(x_a, \bfz_{\perp a})=
\Delta^N D_{h/\aup}(z, \bfz_{\perp h})=0)]$ or integrate to zero,
and neglect it elsewhere.
From Eq. (\ref{sigmaup}), after some straightforward but lengthy algebra, 
one then obtains, for $\pup p \to \pi X$: 
\barr
& & d\sigma^{\uparrow} - d\sigma^{\downarrow} = 
\sum_{a,b,c,d} \int \frac{dx_a \, dx_b}{\pi z} \nonumber \times \\ 
& & \Bigg\{ \int d^2 \bfk_{\perp} \,
\Delta^Nf_{a/\pup} (x_a, \bfk_{\perp}) \> f_{b/p}(x_b) \,
\frac{d \hat \sigma^{ab \to cd}} {d\hat t} (x_a, x_b, \bfk_{\perp}) \>
D_{\pi/c} (z) \label{gen} \\  
& & + \int d^2 \bfk'_{\perp}\,P^{a/\pup}\,f_{a/p}(x_a) \> f_{b/p}(x_b) \>
\Delta_{NN} \hat\sigma^{ab \to cd}(x_a, x_b, \bfk'_\perp) \>
\Delta^N D_{\pi/c}(z, \bfk'_\perp) \Bigg\} \>. \nonumber
\earr
Eq. (\ref{gen}) is worth some comments.

\begin{itemize}
\item
The second line of Eq. (\ref{gen}) represents the Sivers effect:
this expression of $d\sigma^\uparrow - d\sigma^\downarrow$ is the one 
used in Refs. \cite{noi1} and \cite{noi2} to fit the data \cite{e704} 
on $A_N$, Eq. (\ref{an}), and to obtain an explicit functional form for 
$\Delta^Nf_{u/p}$ and $\Delta^Nf_{d/p}$, Eq. (\ref{del1}).

\item
The third line of Eq. (\ref{gen}) represents the Collins effect:
this expression of $d\sigma^\uparrow - d\sigma^\downarrow$ is the one
we are going to use in the sequel to fit the data on $A_N$ and to obtain
an explicit expression for some $\Delta^N D_{\pi/c}$, defined 
in Eq. (\ref{delf1}). Notice that
\beq
\Delta^N D_{\pi/c}(z,\bfk '_{\perp}) = 2\,{\rm Im} \, \hat D_{\pi/c} ^{+-} \;,
\eeq
where [see Eq. (\ref{framp})]
\beq
\hat D_{\pi/c} ^{+-} = \> \sumint_{X, \la_{X}}
\hat{\cal D}_{\la^{\,}_{X}; +} \, \hat{\cal D}^*_{\la^{\,}_{X}; -}
\eeq
is a purely imaginary quantity, by parity invariance.

\item
$P^{a/\pup} = 2i\,\rho_{+,-}^{a/\pup}$ is the polarization of parton $a$ inside 
the transversely polarized proton $\pup$; it is twice the average value of 
the $\uparrow$ component of the parton spin. In this part only collinear 
configurations are taken into account. $P^{a/\pup}$ might depend 
on $x_a$, but we will assume that, at least for large $x_a$ valence quarks, 
it is constant. The product $P^{a/\pup} f_{a/p}(x_a)$ is the transversity 
distribution, often denoted by $\Delta_T q(x_a)$ or $h_1(x_a)$ \cite{jaf}.

\item
$\Delta _{NN} \hat \sigma  (x_a,x_b,\bfk'_{\perp})$ is a 
double spin asymmetry for the the elementary process:
\barr
& & \Delta _{NN} \hat \sigma  (x_a,x_b,\bfk ' _{\perp}) = \left[ 
\frac{ d \hat \sigma ^{a^{\uparrow}b \to c^{\uparrow} d }}{d \hat t} -
\frac{ d \hat \sigma ^{a^{\uparrow}b \to c^{\downarrow} d}}{d \hat t}
\right] = \nonumber \\ & &
\frac{1}{32\,\pi\,\hat s ^2} \, \sum _{\lambda _b ,\lambda _d} 
\left[\hat M_{+, \la^{\,}_d; +, \la^{\,}_b} \,
\hat M^*_{-, \la^{\,}_d; - , \la^{\,}_b} \, -
\hat M_{+, \la^{\,}_d; -, \la^{\,}_b} \,
\hat M^*_{-, \la^{\,}_d; + , \la^{\,}_b} \,\right] \>. \label{dnn}
\earr
   
\item
The division of Eq. (\ref{gen}) into two separate pieces, corresponding 
respectively to spin and $\bfk_\perp$ effects in distribution and 
fragmentation functions, is not due to keeping only leading terms in 
$k_\perp$; the structure of Eq. (\ref{gen}) -- within its range of validity -- 
remains unchanged also at higher order in $k_\perp$, and one should 
only add the appropriate $\bfk_\perp$ dependences in all terms. 
\end{itemize}

\goodbreak
\vskip 12pt
\nd
{\bf 3. Single spin asymmetries in \mbox{\boldmath{$p^{\uparrow} \!\!p \to 
\pi X$}} and Collins effect}
\vskip 6pt

In Refs. \cite{noi1} and \cite{noi2}
we have considered $\Delta^Nf$ as the only possible
source of single spin asymmetries in $\pup p \to \pi X$ processes and
have obtained an explicit expression for it, by fitting the experimental
data. This expression of $\Delta^Nf$ has been used in Ref. \cite{noi2}
to compute the values of $A_N$ for $\bar p^\uparrow p \to \pi X$, in 
agreement with data, and to predict $A_N$ for $\pup p \to KX$, not yet
measured. 

We consider here $\Delta^ND$ as the only possible source of single spin 
asymmetries and see if we can get an equally good description of the data,
obtaining an explicit expression for it, to be used in other processes. 
We will find that this is possible, with some difficulties,
and shall comment on it at the end of this Section.

Of course, both effects -- in distribution and fragmentation functions -- 
might be at work at the same time and be in different proportions
responsible for the observed asymmetries. By considering only one 
effect at a time we can obtain the maximum possible values for 
$\Delta^Nf$ and $\Delta^ND$; such functions could then be used in
processes where only one of them can be at work to produce single
spin asymmetries, and see whether or not they give results in agreement 
with data. For example, in prompt photon production,
$A^\uparrow B \to \gamma X$ \cite{noi2} and Drell-Yan processes, 
$A^\uparrow B \to \mu^+ \mu^- X$,
only $\Delta^Nf$ can be active, whereas in semi-inclusive DIS,
$\ell \pup \to \ell \pi X$, only $\Delta^ND$ should contribute,
because initial state interactions in lepton-proton interactions
are suppressed by powers of $\alpha_{em}$ \cite{alm}. More data are
expected in the future, from running or planned experiments at HERA,
Jefferson Lab and RHIC; a richer and more precise amount of 
experimental information might also allow a simultaneous determination of 
both $\Delta^Nf$ and $\Delta^ND$ by using the full equation (\ref{gen}).      

There is another reason to explore here the effects of $\Delta^ND$ alone.
As we said, the function $\Delta^Nf$ is bound to be zero if the incoming
hadrons are treated as free plane wave states \cite{col, bmt}; this 
conclusion may be avoided by invoking soft initial state interactions,
different from those taken into account in the proof of the  
factorization theorem, together with the assumption that one can still 
use the factorized form (\ref{sigmaup}) (see also Ref. \cite{dra}). 
The same does not apply to the fragmentation
process, which requires anyway final state interactions; it might 
be -- and this has to be determined experimentally -- that the Collins
effect is the main origin of single spin asymmetries. 

We consider then the last line only of Eq. (\ref{gen}) to compute $A_N$,
Eq. (\ref{an}), for $\pup p \to \pi^{\pm,0} X$ processes:
\barr
d\sigma^{\uparrow} - d\sigma^{\downarrow} &=& 
\sum_{a,b,c,d} \int \frac{dx_a \, dx_b}{\pi z} \, d^2 \bfk_{\perp}
\label{dscol} \\ 
&\times& P^{a/\pup}\,f_{a/\pup}(x_a) \> f_{b/p}(x_b) \>
\Delta_{NN} \hat\sigma^{ab \to cd}(x_a, x_b, \bfk_\perp) \>
\Delta^N D_{\pi/c}(z, \bfk_\perp) \nonumber
\earr
and
\barr
d\sigma^{\uparrow} + d\sigma^{\downarrow} &=& 2\, d\sigma^{unp} = 
2 \sum_{a,b,c,d} \int \frac{dx_a \, dx_b}{\pi z} 
\label{dsunp} \\ 
&\times& f_{a/p}(x_a) \> f_{b/p}(x_b) \>
\frac{d\hat\sigma^{ab \to cd}}{d\hat t}(x_a, x_b) \> D_{\pi/c}(z) \>.\nonumber
\earr

Let us consider Eq. (\ref{dscol}) and notice that 
$\Delta^ND_{\pi/c}(z,\bfk_{\perp})$, Eq. (\ref{delf1}), 
is an odd function of $\bfk_{\perp}$: this means that we cannot
neglect the $\bfk_{\perp}$ dependence of the other terms, 
and have to keep into account $\bfk_{\perp}$ values in 
$\Delta_{NN} \hat\sigma^{ab \to cd}$. Eq. (\ref{dscol}) 
can then be written as
\barr
d\sigma^{\uparrow} - d\sigma^{\downarrow} &=& 
\sum_{a,b,c,d} \int \frac {dx_a \, dx_b} {\pi z} \,
d^2\bfk_{\perp} \, P^{a/\pup} \, f_{a/\pup}(x_a) \, f_{b/p}(x_b) \,
\Delta^ND_{\pi/c}(z,\bfk_{\perp}) \nonumber\\
&\times&  \left[ \Delta_{NN}\hat\sigma(x_a, x_b, \bfk_{\perp}) -
\Delta_{NN}\hat\sigma(x_a, x_b, -\bfk_{\perp}) \right] 
\label{ancol}
\earr
where now the integration on $\bfk_{\perp}$ runs only over
one half-plane of its components. 
The only unknown functions in Eq. (\ref{ancol}) are $P^{a/\pup}$ and 
$\Delta^ND_{\pi/c}(z,\bfk_{\perp})$, which, at least in principle, 
are then measurable via the single spin asymmetry $A_N$.

In order to perform numerical calculations, we make some
further assumptions, following Refs. \cite{noi1} and \cite{noi2}.
Our first assumption is that the dominant effect is given by the valence 
quarks in the polarized protons. That is, we assume  
$P^{a/\pup}$ to be non-zero only for valence $u$ and $d$ quarks: 
it is natural to assume a smaller polarization for sea quarks which,
moreover, do not contribute much to the production of large $x_F$ pions. 
Secondly, we evaluate Eq. (\ref{ancol}) by assuming that the main 
contribution comes from $\bfk_{\perp} = \bfk^{0}_{\perp}$ where,
as suggested by Eq. (\ref{sinf}), $\bfk^{0}_{\perp}$ lies in the overall 
scattering plane and its magnitude equals the average value of 
$\langle \bfk^{2}_{\perp}\rangle^{1/2}= k^{0}_{\perp}(z)$. 
This average value does in general depend on $z$. 
The $z$ dependence of the transverse momentum 
of charged pions inside jets was measured at LEP \cite{abr}. 
We have made a fit to their data points. The result is shown 
in Fig. 1 and corresponds to the curve
\beq
\frac{k_{\perp}^0(z)}{M} = 0.61 \; z^{0.27} \; (1-z)^{0.20}\;,
\label{kfit}
\eeq
with $M$ = 1 GeV/$c^2$.

The residual $z$ dependence in $\Delta^ND_{\pi/c}$ not coming from
$k_{\perp}^0$ is taken as a simple power behaviour, so that
\beq
\Delta^ND(z, \bfk_\perp^0) = \frac{k_{\perp}^0(z)}{M} \,
N_c \, z^{\alpha_c}\,(1-z)^{\beta_c} \,,
\label{ix}
\eeq
where $N_c$, $\alpha_c$ and $\beta_c$ are free parameters.

The $\bfk_\perp$ integration then produces the simple expression:
\barr
&&\int d^2\bfk_{\perp} \, \Delta^ND_{\pi/c}(z,\bfk_{\perp}) 
 \left[ \Delta_{NN}\hat\sigma^{ab\to cd} (\bfk_{\perp}) -
 \Delta_{NN}\hat\sigma^{ab\to cd}(-\bfk_{\perp}) \right] \nonumber \\
&\simeq& \frac{k_{\perp}^0(z)}{M} \> N_c \, z^{\alpha_c}(1-z)^{\beta_c}
 \left[ \Delta_{NN}\hat\sigma^{ab\to cd}(\bfk_{\perp}^0) -
 \Delta_{NN}\hat\sigma^{ab\to cd}(-\bfk_{\perp}^0) \right] \>,
\label{app}
\earr
where $k_{\perp}^0(z)$ is given by Eq. (\ref{kfit}).

Exploiting isospin and charge conjugation invariance 
and assuming that a large $z$ pion is mainly 
generated by the fragmentation of a quark which can be a
valence quark for the pion, one can express all functions (\ref{ix}) --
one for each flavour $c$ -- in terms of a single one,
\beq
\Delta^ND_{val}(z, \bfk_\perp^0) = N\, \frac{k_{\perp}^0(z)}{M} 
z^{\alpha}\,(1-z)^{\beta} \,.
\label{ixv}
\eeq
In fact
\barr
& &\Delta^N D_{\pi^+/u}(z, k_{\perp}^0) =
\Delta^N D_{\pi^+/\bar d}(z, k_{\perp}^0) \nonumber \\
&=& \Delta^N D_{\pi^-/\bar u}(z, k_{\perp}^0) =
\Delta^N D_{\pi^-/d}(z, k_{\perp}^0) = \Delta^ND_{val}(z, k_\perp^0)
\label{isopm} 
\earr
and
\barr
& &\Delta^N D_{\pi^0/u}(z, k_{\perp}^0) =
\Delta^N D_{\pi^0/d}(z, k_{\perp}^0)\nonumber \\ 
&=& \Delta^N D_{\pi^0/\bar u}(z, k_{\perp}^0) = 
\Delta^N D_{\pi^0/\bar d}(z, k_{\perp}^0) = 
\frac{1}{2} \; \Delta^N D_{val}(z, k_{\perp}^0) \label{iso0} \>.
\earr

Notice that $\Delta^N D_{\pi/c}$ can be different from zero only for 
polarized quarks, {\it i.e.} quarks resulting from processes $ab \to cd$
for which $\Delta_{NN}\hat\sigma$ is not zero. These are the processes:
$qq \to qq$, $qq' \to qq'$, $q\bar q \to q\bar q$, $q \bar q \to \bar qq$
and $qg \to qg$, and the corresponding $\Delta_{NN}\hat\sigma$ can be 
computed from Eq. (\ref{dnn}). 

As initial polarized partons, like we said before, we consider only
$u$ and $d$ valence quarks. The value of $P^{q/\pup}$ is taken as a free
parameter: given the overall normalization factor $N$ contained in
$\Delta^N D_{val}$ only the relative value 
\beq
P_{u/d} \equiv P^{u/\pup}/P^{d/\pup} \label{prel}
\eeq
is a meaningful free parameter. $P^{u/\pup}$ is assumed to be 2/3. 

In the unpolarized cross-section we take into account all leading order
QCD processes involving quarks and gluons. 
Expressions for the unpolarized partonic distributions and fragmentations 
can be found in the literature; we have chosen, as in Ref. \cite{noi2},
the MRSG parametrization \cite{mrsg} for the partonic distribution
functions, and the parametrization (BKK1) of Ref. \cite{bkk1} for quark 
fragmentation functions into pions. These fragmentations functions are 
sufficiently well established and constrained by data; other sets
available in the  literature have similar large $z$ behaviours and
do not change significantly the quality of the fit and the conclusions
which follow. 

Eqs. (\ref{dscol}) and (\ref{dsunp}) contain now known functions -- the 
unpolarized distribution and fragmentation functions -- computable elementary
dynamics -- $d\hat\sigma/d\hat t$ and $\Delta_{NN}\hat\sigma$ -- 
and unknown functions -- $\Delta^ND_{\pi/c}$ and $P^{a/\pup}$ --
parametrized in a simple way, Eqs. (\ref{kfit}), (\ref{ixv})-(\ref{iso0}) 
and (\ref{prel}),
in terms of the four parameters $N$, $\alpha$, $\beta$ and $P_{u/d}$.

We have computed $A_N$, Eq. (\ref{an}), as a function of $x_F$
in terms of these parameters and have compared with the existing data 
for $\pup p \to \pi^{\pm,0}$ processes \cite{e704}; in our computation
the $p_T$ of the produced pion in the $p\,p$ c.m. frame has been
fixed to $p_T = 1.5$ GeV/$c$, the average transverse momentum of the data.   
We obtain that the $\Delta^N D_{val}$ function which 
best fits the data is given by Eqs. (\ref{kfit}) and (\ref{ixv}) with:
\barr
&&N = - 0.22  \nonumber \\
&&\alpha = 2.33 \qquad\qquad {\rm for} \quad z \le 0.97742 \label{bfit} \\
&&\beta = 0.24 \nonumber 
\earr
and 
\beq
\Delta^ND_{val}(z, k_\perp^0) = - 2 D_{val}(z) = - 2 \times 1.102 \, z^{-1} 
(1-z)^{1.2} \quad \quad {\rm for} \quad z > 0.97742 \>, \label{bfit2}
\eeq
where $z = 0.97742$ is the value at which $|\Delta^ND_{val}(z, k_\perp^0)| = 
2 D_{val}(z)$. $D_{val}(z)$ is taken from Ref. \cite{bkk1}.
The best fit value of $P_{u/d}$ turns out to be
\beq
P_{u/d} = -0.76 \>.
\label{pud}
\eeq

The corresponding fit is shown in Fig. 2.
The fit is qualitatively good and satisfactory --  
one should not forget that we are considering an extreme situation, with 
only Collins effect taken into account and a simple $\bfk_\perp$ dependence 
in $\Delta^ND(z, \bfk_\perp)$; however, the last points at large $x_F$ values 
seem difficult to fit. In our computation $A_N \to 0$ when $x_F \to 1$,
due to the fact that $k_\perp^0(z) \to 0$ when $z \to 1$, Eq. (\ref{kfit}).  
We can obtain the best results only by letting 
$\Delta^N D(z, k_\perp^0)$ to be as large as possible when $z \to 1$,
{\it i.e.} we have to saturate the necessary inequality:
\beq
|\Delta^N D(z, k^0_\perp)| \le 2 D(z) \>. \label{ineq}  
\eeq

In Fig. 3 we plot the ratio $\Delta^N D_{val}(z, k_\perp^0)/2D_{val}$; it shows 
the elementary left-right asymmetry, in the fragmentation of a transversely
polarized valence quark,
\beq
\hat A^N_{\pi/q} \equiv
\frac{\hat D_{\pi/q^\uparrow}(z, k_\perp^0) - 
\hat D_{\pi/q^\downarrow}(z, k_\perp^0)}
{\hat D_{\pi/q^\uparrow}(z, k_\perp^0) + 
\hat D_{\pi/q^\downarrow}(z, k_\perp^0)} \>,
\label{anq}
\eeq
which is needed to fit the data. 

$|\hat A^N_{\pi/q}|$ has to reach the maximum value 1 at large $z$; smaller 
values would give too small values of $A_N$ at large $x_F$.

This was not the case when relying only on Sivers effect \cite{noi1, noi2}. 
In Fig. 4 we plot the value of the left-right asymmetry in the splitting
of a transversely polarized proton into a valence quark, 
\beq
\hat A^N_{q/p} \equiv
\frac{\hat f_{q/p^\uparrow}(x, k_\perp^0) - 
\hat f_{q/p^\downarrow}(x, k_\perp^0)}
{\hat f_{q/p^\uparrow}(x, k_\perp^0) + 
\hat f_{q/p^\downarrow}(z, k_\perp^0)} 
= \frac{\Delta^N f_{q/p}(x, k_\perp^0)}{2f_{q/p}} \>, 
\label{anp}
\eeq
as obtained from Refs. \cite{noi2} for $u$ and $d$ valence 
quarks\footnote{Notice that, due to a mistake in the writing of the paper, 
the values of $N_a$ given in Eq. (9) of Ref. \cite{noi2} are wrong and 
have to be multiplied by a factor 4; the fit to the data, Fig. 1 of  
Ref. \cite{noi2}, remains the same.}.

By comparing Figs. 3 and 4 we see that a small elementary left-right 
asymmetry $\hat A^N_{u,d/p}$ in the proton distribution alone allows to fit 
the data on $A_N$ \cite{e704, noi2}; instead, one needs a much bigger (at 
large $z$) left-right asymmetry $\hat A^N_{\pi/q}$ in the quark fragmentation 
alone in order to fit the same data. Moreover, the fit obtained here with 
the large $\hat A^N_{\pi/q}$ might show some difficulty at the largest 
$x_F$ values. If such a discrepancy should be confirmed by
further data it would be a significative indication towards the importance 
of spin and $\bfk_\perp$ dependences in distribution functions. 

We also notice that in the simple model proposed by Collins for $\Delta^ND$
\cite{col}, one finds that $\hat A^N_{\pi/q} \to 0$ when $z \to 1$; 
such model could not explain the observed values of $A_N$. 

The reason for the different results due solely to $\Delta^Nf$ 
\cite{noi1, noi2} or $\Delta^ND$ (this paper) can be qualitatively
understood by looking again at Eq. (\ref{gen}) and it turns out to be very
interesting, as it is due to the different dynamical contributions
in the two cases: let us compare Eq. (\ref{app}) with its analogue for 
Sivers contribution, obtainable from the second line of Eq. (\ref{gen})
\cite{noi2}:
\barr
&&\int d^2\bfk_{\perp} \, \Delta^Nf_{a/\pup}(x_a,\bfk_{\perp}) 
 \left[ \frac {d\hat\sigma^{ab\to cd}}{d\hat t} (\bfk_{\perp}) -
        \frac {d\hat\sigma^{ab\to cd}}{d\hat t}(-\bfk_{\perp}) \right] 
\nonumber \\
&\simeq& \frac{k_{\perp}^0(x_a)}{M} \> N_a \, z^{\alpha_a}(1-z)^{\beta_a}
 \left[ \frac {d\hat\sigma^{ab\to cd}}{d\hat t} (\bfk_{\perp}^0) -
        \frac {d\hat\sigma^{ab\to cd}}{d\hat t} (-\bfk_{\perp}^0) \right] \>.
\label{apps}
\earr
In the expression for $A_N$ Eqs. (\ref{app}) or (\ref{apps}) have to 
be divided by the unpolarized cross section which contains $f$, $D$ and
$d\hat\sigma/d\hat t$. Now, it turns out that not only $\Delta_{NN} \hat\sigma$
is smaller than $d\hat\sigma/d\hat t$, but also the cancellation between the 
two terms in squared brackets is stronger for $\Delta_{NN} \hat\sigma$, 
Eq. (\ref{app}), than for $d\hat\sigma/d\hat t$, Eq. (\ref{apps}).
This explains why fitting the data requires a much bigger value of 
$|\Delta^ND|/2D$ than $|\Delta^Nf|/2f$.  

We finally comment on the value found for $P_{u/d}$ which might be 
surprising: in $SU(6)$, for example,
one has $P^{u/\pup} = 2/3$ and $P^{d/\pup} = -1/3$, so that $P_{u/d} = -2$.
We find here that, indeed, the $u$ and $d$ valence quark polarizations 
inside a polarized proton have opposite signs, but the $d$ polarization
is, in magnitude, slightly bigger than the $u$ polarization, Eq. (\ref{pud}).

\goodbreak
\vskip 12pt
\nd
{\bf 4. Application to other processes, \mbox{\boldmath{$\bar p^{\uparrow} 
\!\!p \to \pi X$}} and \mbox{\boldmath{$p^{\uparrow} \!\! p \to KX$}}}
\vskip 6pt
We now apply the $\Delta^N D_{val}(z,\bfk_{\perp}^0)$ we have obtained 
by fitting the E704 experimental data on pion production, to predict single 
spin asymmetries in other processes.

Recently the E704 Collaboration at Fermilab has presented results on single 
spin asymmetries for inclusive production of pions also in the collision of 
transversely polarized antiprotons off a proton target \cite{e7042}. 
The kinematical conditions are the same as in the case of polarized 
protons \cite{e704}. Within our model the connection between single spin
asymmetries with polarized protons or antiprotons is very simple: 
since only valence quark contributions are taken into account, we
just have to exploit the charge conjugation relations (\ref{isopm}) and
(\ref{iso0}) for the $\Delta^ND$ and similar ones for the 
$P^{\bar q/\bar p^\uparrow}$. In particular, this means that we should expect
\beq
A_N(\bar{p}^\uparrow p\to\pi^\pm) \simeq A_N(\pup p\to\pi^\mp)
\quad\quad\quad A_N(\bar{p}^\uparrow p\to\pi^0)
\simeq A_N(\pup p\to\pi^0) \,.
\label{resppbar}
\eeq

Our results are shown in Fig. 5, and indeed they agree with
Eq. (\ref{resppbar}). The experimental data are in very good agreement
with our predictions for $\pi^+$ and somewhat smaller than our predictions
for $\pi^0$ and $\pi^-$, although the overall agreement is good. 
The expectations (\ref{resppbar}) are essentially based on charge conjugation 
invariance and the dominance of certain elementary dynamical contributions
and it is difficult to see how they could not be true; we expect that
further data both on $\pup p \to \pi X$ and $\bar{p}^\uparrow p \to \pi X$
should lead to a better agreement with Eq. (\ref{resppbar}).

We consider now the case where charged and neutral kaons are detected in the 
final state, $\pup p \to KX$; the kinematical conditions are the same as
for the other processes. Lacking a determination of 
$\Delta^ND(z, \bfk_\perp^0)$ for the
fragmentation of quarks into kaons we assume that the
$\Delta^ND_{val}$ are the same for $K$ and $\pi$ production, up to a 
different normalization factor which we fix by imposing, for valence quarks,
\beq
\frac{\Delta^ND_{K/q}(z, \bfk_\perp^0)}{\Delta^ND_{\pi/q}(z, \bfk_\perp^0)}
= \frac{\langle N_K \rangle}{\langle N_\pi \rangle} \,,
\label{risc}
\eeq
where $\langle N_h \rangle = \int_0^1 dz \, D_{h/q}(z)$. We have used 
unpolarized kaon fragmentation function sets from Refs. 
\cite{bkk1, bkk2, gr, imr} and have computed the appropriate normalization 
factor $\langle N_K \rangle / \langle N_\pi \rangle$ for each set.

We should keep in mind that here only large $x$ {\it valence} quarks 
from the initial polarized proton are taken into account: this means that
contribution from strange quarks from the polarized proton sea are completely 
neglected and the production of kaons is through non-strange quarks only. 
The eventual role of strange polarized quarks (and antiquarks) could only 
be studied within a more refined model for the transversely polarized 
distribution functions, $h_1(x)=P^{q/p^{\uparrow}} \, f_{q/p}(x)$, 
where contributions from $q=s$ and $\bar s$ are included. However,
we have checked that, even assuming a large value of $P^{s/\pup}$
and taking $\Delta^N D_{K/s} = \Delta^N D_{val}$, the contribution
of strange quarks to $A_N$ is negligible at large $x_F$, 
as $f_{s/p}(x_a)$ is small at large $x_a$ values.   
 
Fig. 6 shows the results we obtain for $A_N$ in  
the production of either charged or neutral 
kaons, $(K^+ + K^-)/2$ and $(K^0 + \bar K^0)/2$ respectively.
The positive values of $A_N$ refer to charged kaons, while the negative
ones to neutral ones: the values obtained are large for all sets of 
unpolarized fragmentation functions, whose features are 
further discussed in Ref. \cite{noi2}. Notice that the different 
sets of fragmentation functions give qualitatively similar results,
contrary to the findings of Refs. \cite{noi2}, where, for neutral kaons,
the results show a strong dependence on the choice of the fragmentation
functions. 

Let us stress once more that these results for $K$ production are 
rather qualitative and based on the simple assumption about the form
of $\Delta^ND_{K/q}(z, \bfk_\perp^0)$, Eq. (\ref{risc}); 
nevertheless, they clearly show how $A_N$ might be
large and measurable also for $\pup p \to KX$ processes and its 
measurement should give more information about the Collins effects
in the fragmentation of a polarized quark into a kaon.

\goodbreak
\vskip 12pt
\nd
{\bf  5 Comments and conclusions}
\vskip 6pt
 
Single spin asymmetries in inclusive hadron production can be explained
within QCD; they are twist-3 effects which can be large and measurable
in the $p_T$ range of few GeV/$c$. Several possible origins -- leading
to the introduction of new basic functions -- have been
discussed in the literature, but a clear and comprehensive understanding
has not been reached yet: more theoretical and experimental work 
is still necessary. 

In this paper and previously in Refs. \cite{noi1} and \cite{noi2} we attempt
a consistent phenomenological description and computation of single spin
asymmetries based on QCD parton model and the introduction of fundamental
spin asymmetries at the quark level, both in the splitting of polarized
protons into quarks, Eq. (\ref{del1}), and in the fragmentation of
polarized quarks into hadrons, Eq. (\ref{delf1}). The link between
this approach and other approaches based on the introduction of parton
correlation functions \cite{qs3, rat} and/or gluonic poles \cite{bmt}
has to be clarified. As suggested by Ratcliffe \cite{rat} the data might 
also be analysed using a general parametrization of the partonic
correlators. 

In Ref. \cite{noi1} we showed how the so-called Sivers effect alone could
account for data on $A_N$ and we derived an explicit expression for
the function $\Delta^Nf$; here we have shown that also the Collins effect 
-- with some more difficulties -- can account for the same data, 
and we have obtained an explicit expression for the function $\Delta^ND$. 
We have also introduced a general formalism which includes both effects, 
Eq. (\ref{gen}). 

Their relative importance has to be established by further experimental 
data; single spin asymmetries could be measured in the near future 
at Jefferson Lab, HERA \cite{now} and RHIC. Of particular interest are those 
processes in which only one of the two effects is expected to be active: in 
semi-inclusive DIS, $\ell \pup \to \ell hX$, and in $\gamma^* \pup \to hX$ 
initial state interactions are suppressed by higher powers of $\alpha_{em}$
and single spin asymmetries should originate only from the fragmentation
of a final polarized quark \cite{noi4}. It would be very interesting
to have data on such processes. 
 
On the other hand processes like $\pup N \to \gamma X$ and 
$\pup N \to \mu^+\mu^- X$ could exhibit single spin asymmetries only
due to spin and $\bfk_\perp$ dependent distribution functions \cite{noi2}
or quark-gluon correlations \cite{qs1, bmt, sch}; also data on these 
processes would be of great help. 

A comprehensive phenomenological description and a physical understanding 
of subtle single spin effects, within perturbative QCD and a simple  
factorization model, seems indeed possible; more data should allow the 
determination of new basic distribution and fragmentation functions,
allowing genuine predictions for new processes and a deeper knowledge
of the nucleon structure and the hadronization mechanism. More theoretical
work, towards a true generalization of the factorization theorem to include
transverse momenta, is also needed and might be motivated by our 
phenomenological study.  

\vskip 18pt
\noindent
{\bf Acknowledgements}
\vskip 6pt
We are grateful to Piet Mulders for many interesting discussions;
M. B. acknowledges support from the EU-TMR program, contract n. FRMX-T96-0008.


\newpage
\noindent
{\bf Figure Captions}

\vskip 12pt
\noindent
Figure 1. The pion intrinsic average transverse momentum $k_{\perp}^0$ in a 
jet as a function of $z$. The diamonds are the data from Abreu {\it et al.} 
[19], the continuous line is our fit, as given by Eq.~(17). 

\vskip 12pt
\noindent
Figure 2. Single spin asymmetry for pion production, $\pup \, p \,
\to \, \pi \, X$. The data points are the 
E704 experimental single spin asymmetries [3] for $\pi ^+$ (diamonds),  
$\pi ^0$ (squares) and $\pi ^-$ (triangles). The solid line is our best fit 
for $\pi ^+$, the dashed line for $\pi^0$ and the dotted line for $\pi^-$, 
obtained under the assumption of Collins effect only.

\vskip 12pt
\noindent
Figure 3. The ratio between $|\Delta ^N D_{val}|$ and twice the valence 
unpolarized fragmentation function $D$, Eq. (28), as a 
function of $z$. Notice that $|\Delta ^N D| /2D = 1$ for $z \geq 0.97742$.
 
\vskip 12pt
\noindent
Figure 4. The ratio between $|\Delta ^N f _{val}|$ and twice the valence
unpolarized distribution function $f$, Eq. (29), as a 
function of $x$. The solid line refers to $u$ quarks and the dashed line 
to $d$ quarks. Notice that in both cases 
$|\Delta ^N f| /2f \to 0$ for $x \to 1$.
 
\vskip 12pt
\noindent
Figure 5. Single spin asymmetry for pion production in 
$\overline p ^{\uparrow} \, p \, \to \, \pi \, X$. 
The data points are the E704 experimental single spin asymmetries [22] 
for $\pi ^+$ (diamonds), $\pi ^0$ (squares) and $\pi ^-$ (triangles). 
The solid line is the result of our model (with Collins effect only)
for $\pi ^+$, the dashed line for $\pi^0$ and the dotted line for $\pi^-$.

\vskip 12pt
\noindent
Figure 6. Single spin asymmetry, as predicted by our model (with Collins 
effects only), for the production of charged (upper curves) 
and neutral (lower curves) kaons, $(K^+ + K^-)/2$ and 
$(K^0 + \bar K^0)/2$ respectively. The solid lines correspond to the  
BKK1 set of unpolarized fragmentation functions [21], the dashed lines 
correspond to the BKK2 set [23], the dotted lines to the GR set [24], and the 
dashed-dotted lines to the IMR set [25]; see Ref.~[4] and text for further 
details.

\newpage

\begin{figure}[t]
\begin{center}
\mbox{~\epsfig{file=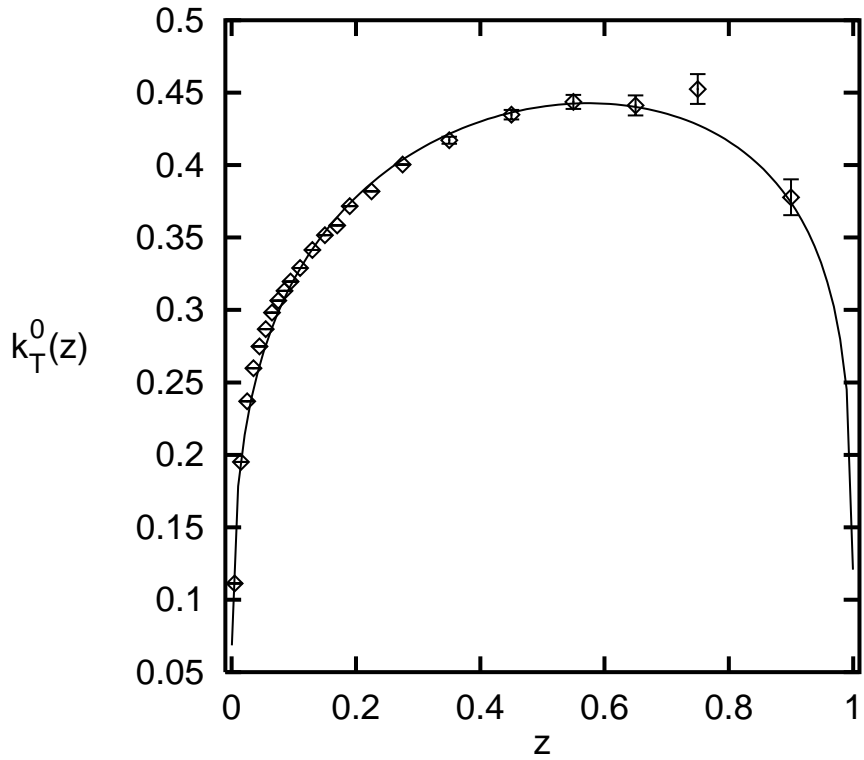,angle=-90,width=12cm}}
\caption{The pion intrinsic average transverse momentum $k_{\perp}^0$ in a jet
as a function of $z$. The diamonds are the data from Abreu {\it et al.} [19], 
the continuous line is our fit, as given by equation (17). }
\end{center}
\end{figure}

\newpage

\begin{figure}[t]
\begin{center}
\mbox{~\epsfig{file=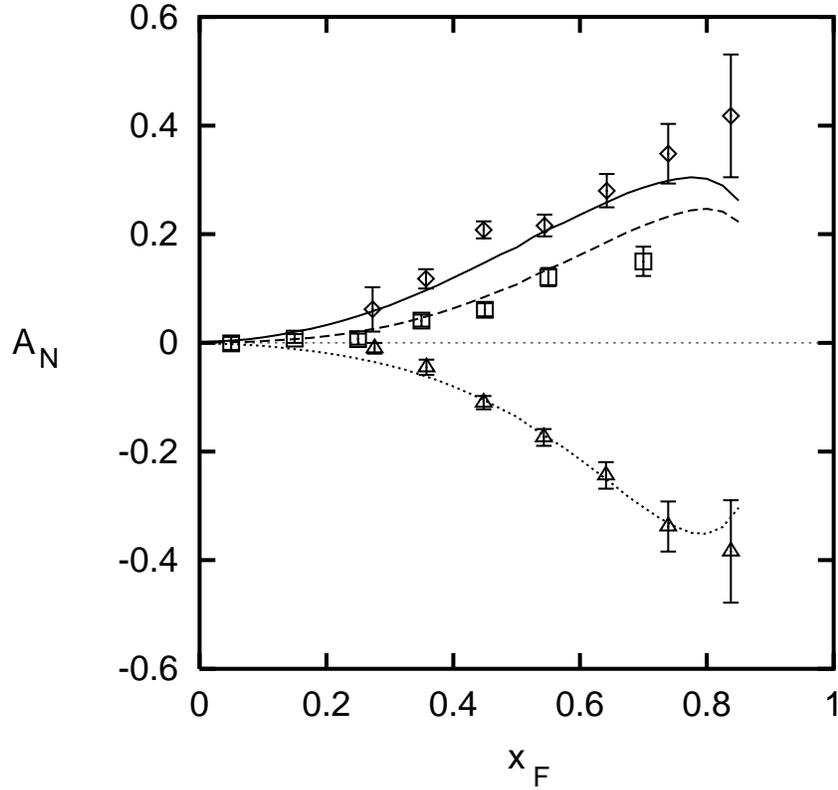,angle=-90,width=12cm}}
\vspace{1cm}
\caption{ 
Single spin asymmetry for pion production, $\pup \, p \to \, \pi \, X$.
The data points are the 
E704 experimental single spin asymmetry for $\pi ^+$ (diamonds),  
$\pi ^0$ (squares) and $\pi ^-$ (triangles) [3]. The solid line is our best 
fit for $\pi ^+$, the dashed line for $\pi^0$ and the dotted line for $\pi^-$, 
obtained under the assumption of Collins effect only.}
\end{center}
\end{figure}

\newpage

\begin{figure}[t]
\begin{center}
\mbox{~\epsfig{file=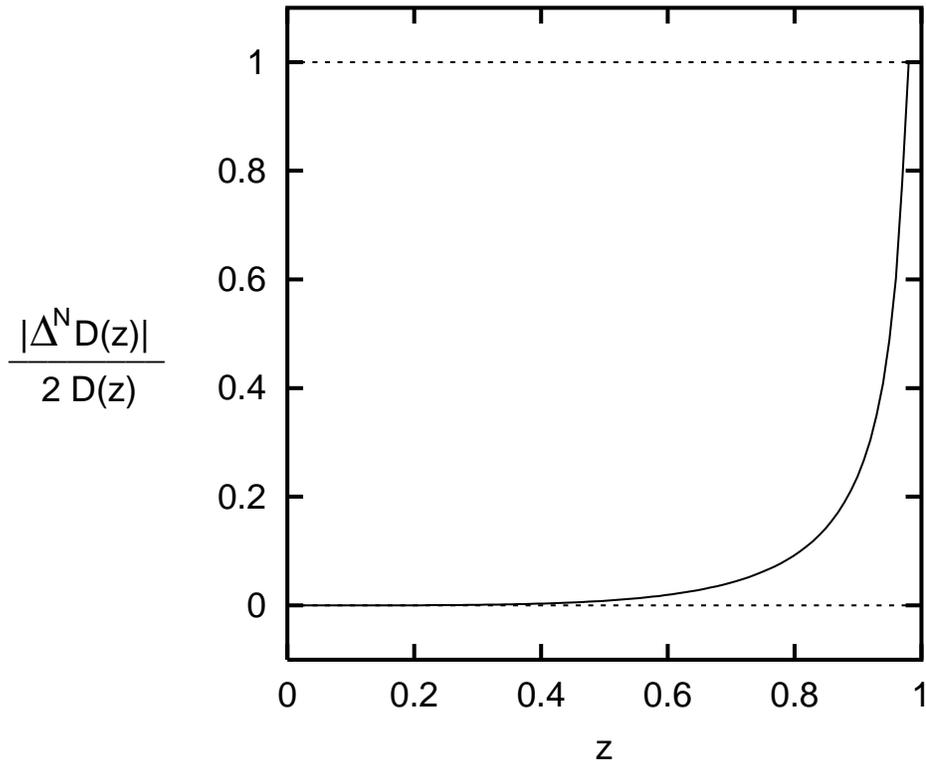,angle=-90,width=12cm}}
\caption{ 
The ratio between $|\Delta ^N D _{val}|$ and twice the valence 
unpolarized fragmentation function $D$, Eq. (28), as a function of $z$. 
Notice that $|\Delta ^N D| /2D = 1$ for $z \geq 0.97742$.}
\end{center}
\end{figure}

\newpage

\begin{figure}[t]
\begin{center}
\mbox{~\epsfig{file=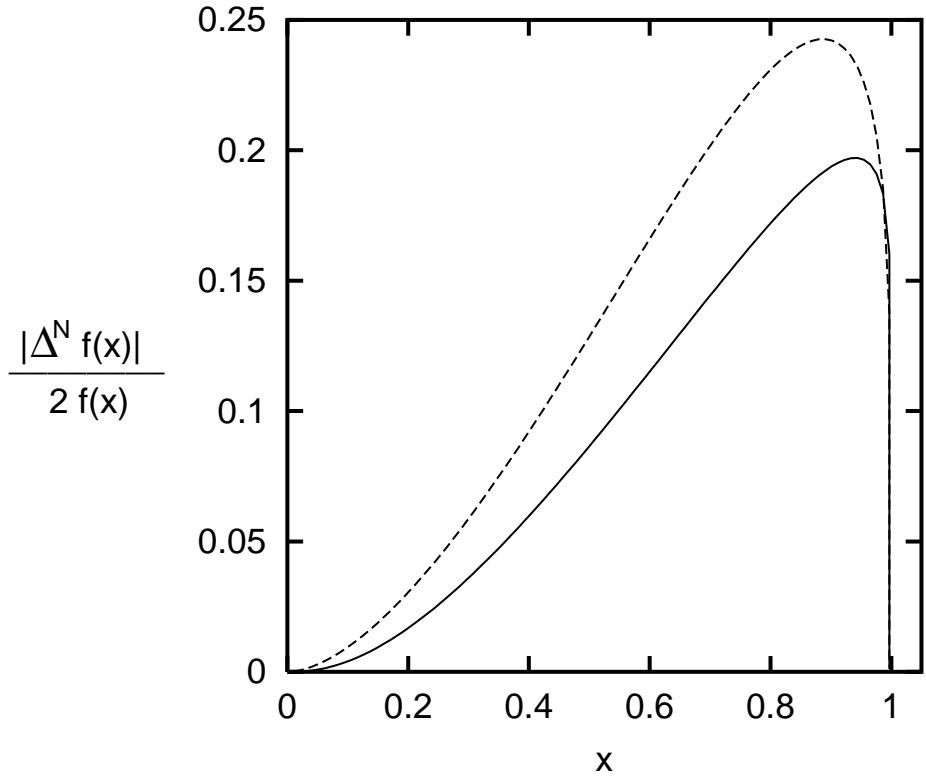,angle=-90,width=12cm}}
\caption{The ratio between $|\Delta ^N f _{val}|$ and twice the valence 
unpolarized distribution function $f$, Eq. (29), as a 
function of $x$. The solid line refers to $u$ quarks and the dashed line 
to $d$ quarks. Notice that in both cases
$|\Delta ^N f| /2f \to 0$ for $x \to 1$.}
\end{center}
\end{figure}

\newpage

\begin{figure}[t]
\begin{center}
\mbox{~\epsfig{file=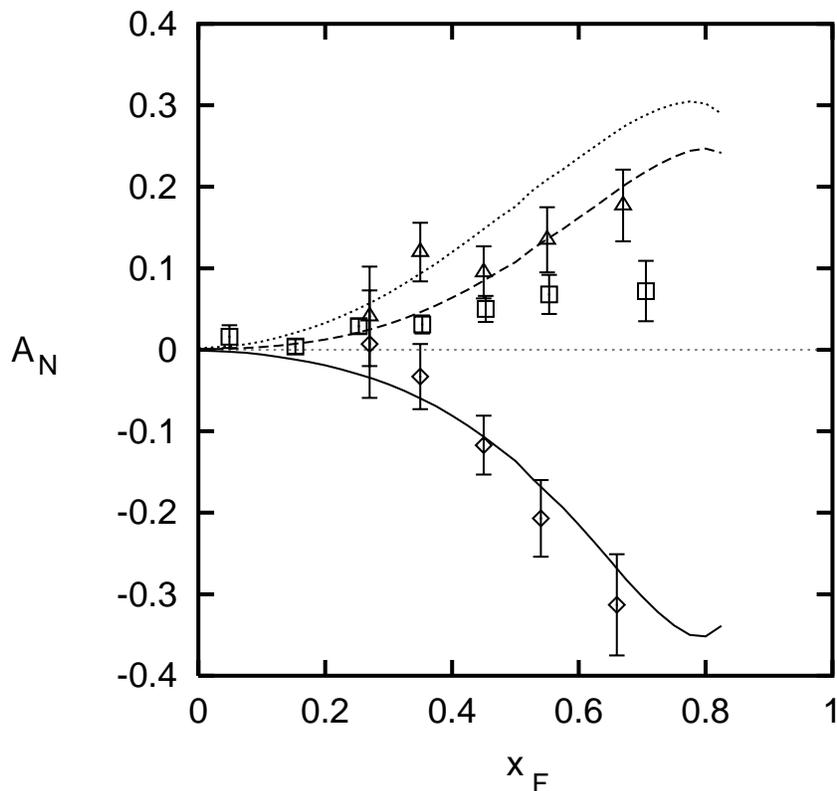,angle=-90,width=12cm}}
\caption{ 
Single spin asymmetry for pion production in 
$\overline p ^{\uparrow} \, p \, \to \, \pi \, X$. 
The data points are the E704 experimental single spin asymmetries [22] 
for $\pi ^+$ (diamonds), $\pi ^0$ (squares) and $\pi ^-$ (triangles). 
The solid line is the result of our model (with Collins effect only)
for $\pi ^+$, the dashed line for $\pi^0$ and the dotted line for $\pi^-$.}
\end{center}
\end{figure}

\newpage

\begin{figure}[t]
\begin{center}
\mbox{~\epsfig{file=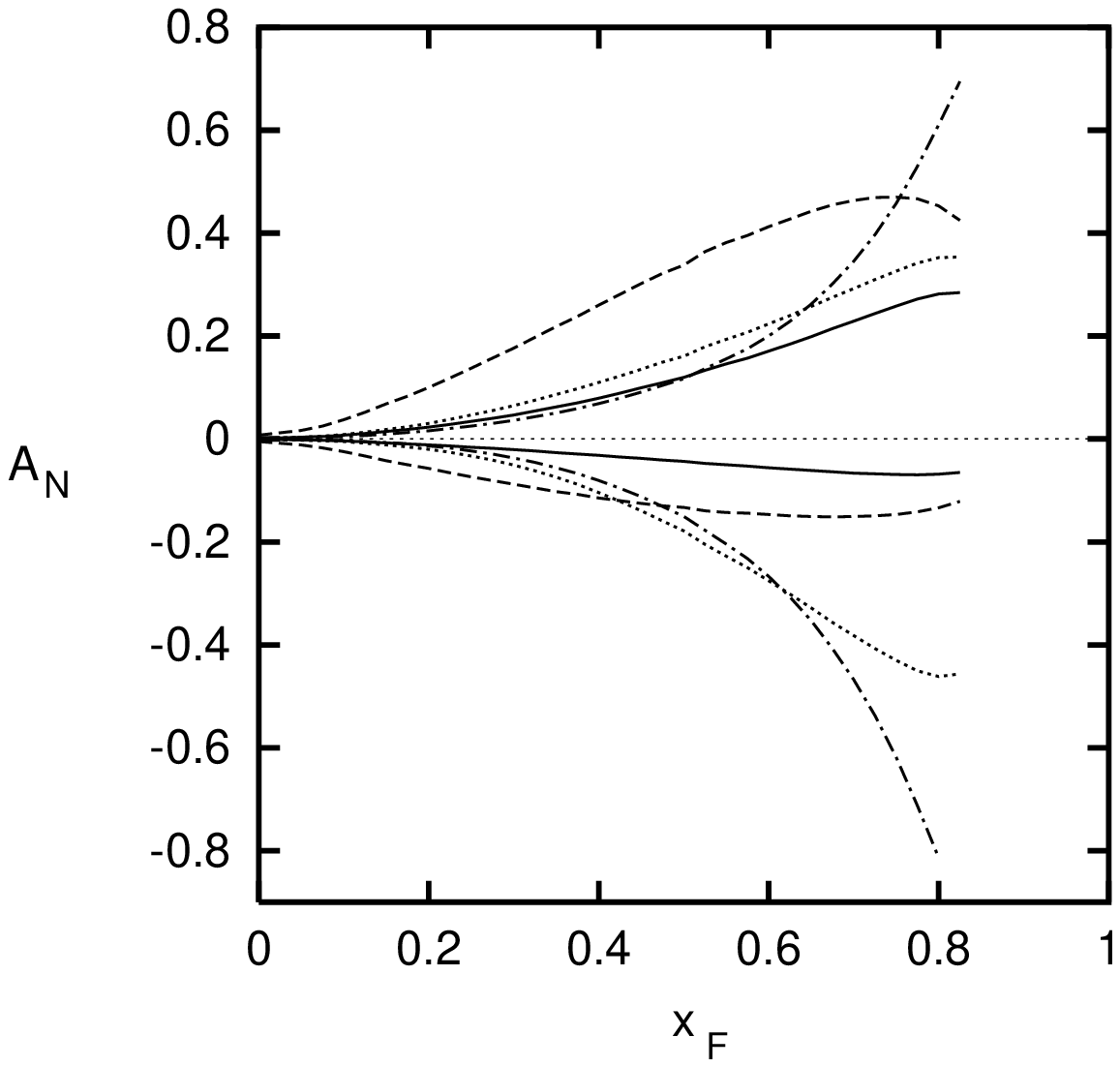,angle=-90,width=12cm}}
\vspace{0.6cm}
\caption{ Single spin asymmetry, as predicted by our model (with Collins 
effects only), for the production of charged (upper curves) 
and neutral (lower curves) kaons, $(K^+ + K^-)/2$ and 
$(K^0 + \bar K^0)/2$ respectively. The solid lines correspond to the  
BKK1 set of unpolarized fragmentation functions [21], the dashed lines 
correspond to the BKK2 set [23], the dotted lines to the GR set [24], and the 
dashed-dotted lines to the IMR set [25]; see Ref.~[4] and text for further 
details.}
\end{center}
\end{figure}


\newpage
\baselineskip=6pt
\small
\begin{thebibliography}{99}
\setlength{\parskip}{6pt}
\bibitem{cms}
\vskip-8pt
J.C. Collins, D.E. Soper and G. Sterman, in {\it Perturbative QCD}, 
Ed. A.H. Mueller, World Scientific, Singapore, 1989
\bibitem{noi1}
\vskip-8pt
M. Anselmino, M. Boglione and F. Murgia, \PL{B362} (1995) 164
\bibitem{e704}
\vskip-8pt
D.L. Adams {\it et al.}, \PL{B264} (1991) 462; \PRL{77} (1996) 2626
\bibitem{noi2}
\vskip-8pt
M. Anselmino and F. Murgia, \PL{B442} (1998) 470, 
e-Print Archive: hep-ph/9808426
\bibitem{siv}
\vskip-8pt
D. Sivers, \PR{D41} (1990) 83; {\bf D41} (1991) 261
\bibitem{col}
\vskip-8pt
J.C. Collins, \NP{B396} (1993) 161
\bibitem{art}
\vskip-8pt
X. Artru, J. Czyzewski and H. Yabuki, \ZP{C73} (1997) 527
\bibitem{qs1}
\vskip-8pt
J.W. Qiu and G. Sterman, \NP{B353} (1991) 137
\bibitem{qs2}
\vskip-8pt
J.W. Qiu and G. Sterman, 
Proceedings of the ``Polarized Collider Workshop", University Park, PA 1990,
J. Collins, S.F. Heppelman and R.W. Robinett Eds., AIP Conference Proceedings
n. 223, {\it Particles and Fields Series} {\bf 42}, p. 249 
\bibitem{qs3}
\vskip-8pt
J.W. Qiu and G. Sterman, \PR{D59} (1999) 014004, 
e-Print Archive: hep-ph/9806356
\bibitem{mul1}
\vskip-8pt
D. Boer and P. Mulders, \PR{D57} (1998) 5780; for a comprehensive review
of $T$-odd distribution functions see D. Boer, PhD thesis. 
\bibitem{bmt}
\vskip-8pt
D. Boer, P.J. Mulders and O. Teryaev, \PR{D57} (1998) 3057 
\bibitem{bm}
\vskip-8pt
M. Boglione and P. Mulders, e-Print Archive: hep-ph/9903354 
\bibitem{mul2}
\vskip-8pt
R. Jacob and P.J. Mulders, Proceedings of SPIN 96, World Scientific 1997,
e-Print Archive: hep-ph/9610295 
\bibitem{noi3}
\vskip-8pt
M. Anselmino, M. Boglione and F. Murgia, Proceedings of Trends in Collider 
Spin Physics, Trieste 1995, World Scientific 1997, p. 194;
e-Print Archive: hep-ph/9604397 
\bibitem{jaf}
\vskip-8pt
R.L. Jaffe and Xiangdong Ji, \NP{B375} (1992) 527
\bibitem{alm}
\vskip-8pt
M. Anselmino, E. Leader and F. Murgia, \PR{D56} (1997) 6021 
\bibitem{dra}
\vskip-8pt
M. Anselmino, A. Drago and F. Murgia, e-Print Archive: hep-ph/9703303 
\bibitem{abr}
\vskip-8pt
DELPHI Collaboration, P. Abreu {\it et al.}, \ZP{C65} (1995) 11
\bibitem{mrsg}
\vskip-8pt
A.D. Martin, W.J. Stirling and R.G. Roberts, \PL{B354} (1995) 155 
\bibitem{bkk1}
\vskip-8pt
J. Binnewies, B.A. Kniehl and G. Kramer, \ZP{C64} (1995) 471
\bibitem{e7042}
\vskip-8pt
A. Bravar {\it et al.}, \PRL{77} (1996) 2626
\bibitem{bkk2}
\vskip-8pt
J. Binnewies, B.A. Kniehl and G. Kramer, 
\PR{D52} (1995) 4947; \PR{D53} (1996) 3573 
\bibitem{gr}
\vskip-8pt
M. Greco, S. Rolli and A. Vicini, \ZP{C65} (1995) 277; 
M. Greco and S. Rolli, \PR{D52} (1995) 3853 
\bibitem{imr}
\vskip-8pt
D. Indumathi, H.S. Mani and A. Rastogi, \PR{D58} (1998) 094014
\bibitem{rat}
\vskip-8pt
P.G. Ratcliffe, e-Print Archive: hep-ph/9806369
\bibitem{now}
\vskip-8pt
M. Anselmino {\it et al.}, DESY report DESY-96-128,
e-Print Archive: hep-ph/9608393 
\bibitem{noi4}
\vskip-8pt
M. Anselmino, M. Boglione and F. Murgia, in preparation
\bibitem{sch}
\vskip-8pt
N. Hammon, B. Ehrnsperger and A. Schaefer, {\it J. Phys.} {\bf G24} (1998) 991 
\end{thebibliography}
\end{document}